# Three related topics on the periodic tables of elements

Yoshiteru Maeno*, Kouichi Hagino, and Takehiko Ishiguro

Department of physics, Kyoto University, Kyoto 606-8502, Japan

* maeno.yoshiteru.2e@kyoto-u.ac.jp

**Abstaract:** A large variety of periodic tables of the chemical elements have been proposed. It was Mendeleev who proposed a periodic table based on the extensive periodic law and predicted a number of unknown elements at that time. The periodic table currently used worldwide is of a long form pioneered by Werner in 1905. As the first topic, we describe the work of Pfeiffer (1920), who refined Werner's work and rearranged the rare-earth elements in a separate table below the main table for convenience. Today's widely used periodic table essentially inherits Pfeiffer's arrangements. Although long-form tables more precisely represent electron orbitals around a nucleus, they lose some of the features of Mendeleev's short-form table to express similarities of chemical properties of elements when forming compounds. As the second topic, we compare various three-dimensional helical periodic tables that resolve some of the shortcomings of the long-form periodic tables in this respect. In particular, we explain how the 3D periodic table *"Elementouch"* (Maeno 2001), which combines the *s*- and *p*-blocks into one tube, can recover features of Mendeleev's periodic law. Finally we introduce a topic on the recently proposed nuclear periodic table based on the proton magic numbers (Hagino and Maeno 2020). Here, the nuclear shell structure leads to a new arrangement of the elements with the proton magic-number nuclei treated like noble-gas atoms. We show that the resulting alignments of the elements in both the atomic and nuclear periodic tables are common over about two thirds of the tables because of a fortuitous coincidence in their magic numbers.

## 1. Introduction

Periodic table of the chemical elements is undoubtedly considered as one of the greatest scientific achievements of humanity. It expresses the periodic properties of the building blocks of nature in a concise table. Mendeleev's periodic table is based on the "periodic law" in which the chemical and physical properties of elements and their compounds are periodic functions of the atomic weight (Mendeleev 1869, 1871). With the discoveries of new classes of elements and accurate determination of their chemical and physical properties, along with the development of quantum mechanics that introduced fundamental concepts in elements, numerous efforts have been made to improve the periodic tables of elements (van Spronsen 1969, Mazurs 1974, Imyanitov 2016, Scerri 2020). In this article, we discuss the following three topics. (1) *Long-form periodic tables by Werner and by Pfeiffer*. For the evolution of the long-form table we use today, Werner's pioneering work in 1905 is known (van Spronsen 1969). However, it is much less known who introduced the arrangement of the rare-earth elements in a separate table below the main table for conciseness and convenience. We introduce the work of Pfeiffer (Pfeiffer 1920), who was a student and an assistant to Werner. (2) *Three dimensional (3D) periodic table that expresses Mendeleev's periodic law*. We will explain how valence tendencies, clearly expressed in Mendeleev's short-period table, are recovered in the 3D helical table based on the modern long form, if the element symbols are arranged on *three* concentric tubes. (3) *Comparison between nuclear and atomic periodic tables of elements.* Protons and neutrons in a nucleus form shell structures of nucleon orbitals, analogous to the shell structure of electron orbitals around a nucleus in an atom. Thus, it is possible to make a nuclear periodic table based on proton magic number nuclei, corresponding to the noble-gas (rare-gas) elements. We describe how both nuclear and atomic periodic tables happen to have common arrangements over many elements.

**2. Pioneering work towards modern long-form periodic tables: Werner (1905) and Pfeiffer (1920)**

Mendeleev's periodic table Tabelle II (Mendeleev 1871) is a short form consisting of eight groups. It is quite different from the long-form table currently used worldwide. Mendeleev also discussed a variation of PTs to separate the representative and transition elements into different rows as early as in 1869 (Mazurs 1974), and later presented a 17-column horizontal table in 1879 (Mendeleev 1879). The long-form table clearly separating the rare-earth elements as yet additional groups was presented by Alfred Werner in 1905 (Werner 1905). Werner's original table (Fig. 1) consists of 32 columns combining the rare-earth elements in the "long cesium period". It should be noted that this table was proposed many years before the Bohr model of the atom (1913). The consistency of the alignment of the long-form table with the quantum-mechanical atomic structure makes this type of tables much better accepted in modern times. Although Werner is a prestigious chemist, widely known as a founder of coordination chemistry and a recipient of the Nobel prize in chemistry in 1913, his contribution to the modern periodic table is not as widely recognized as it should be. The description in Sec. 6.5 of van Spronsen's work (van Spronsen 1969) may give us a hint why his proposal was not as properly recognized. His paper was initially disputed by Augsto Piccini as well as by Richard Abegg, for the reason that by separating the sub-groups from the main groups Werner's table no longer satisfies the main principles of the periodic system. In addition, Werner's table contains the controversy over the incorrect ordering of Nd and Pr (in Sec. 8.6).

Fig. 1 Werner's long-form periodic table (Werner 1905).

Werner's table with 32 columns is not very convenient to use because of its very long structure. The periodic table widely used today is with a modified arrangement in which the rare-earth elements are placed separately below the main table to make it more concise and convenient to use. However, it is much less known who pioneered such modified arrangement (Okuno (Zain-shi) 1974, Robinson 2018). In this article we would like to point out that it was Pfeiffer who proposed the arrangement with 18 columns as shown in Fig. 2 in 1920 (Pfeiffer 1920). Von Paul Pfeiffer (21 April 1875 – 4 March 1951) was an influential German chemist (Oesper 1951). He received his Ph.D. in 1898 at the University of Zurich, studying under Alfred Werner. Pfeiffer was considered as Werner's most successful student and became Werner's assistant and then worked as an associate professor in Zurich until 1916, when he moved to Rostock, to Karlsruhe in 1919, and finally to Bonn in 1922. At Bonn, where he had studied as an undergraduate, he occupied prestigious Kekulé's chair.

Pfeiffer followed Werner's work and extended the table incorporating the knowledge from X-ray physics. His long-form periodic table (Fig. 2) consists of 18 columns and the rare-earth elements are arranged in a separate table. He wrote in his 1920 paper "the version of the periodic system given here has been used for several years in my lectures on inorganic chemistry; it has proven to be very good." Thus he seems to have used this table in his lectures in Rostock and Karlsruhe, and possibly even in Zurich.

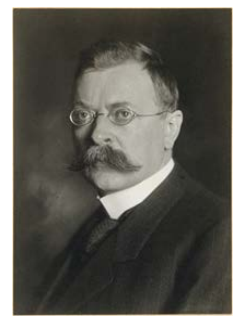

(a) (b)

Fig. 2 (a) Pfeiffer's periodic table (Pfeiffer 1920) along with (b) his portrait in 1914 (Wikipedia).

Pfeiffer's paper in 1920 entitled "the fertilization of chemistry by X-ray physics" consists of two parts. In the first part "the X-ray physics and the valence problems", he discussed the application of newly developed X-ray techniques to the determination of crystal structures of chemical compounds, especially of molecular structures where stereochemical laws apply. In the second part "the X-ray physics and the periodic system of the elements", he reviewed the known elements at the time and six missing elements to be discovered. "Since uranium has the atomic number 92 according to Moseley, we have 92 from hydrogen to uranium and no more elements". Concerning the missing elements, he noted "two rare earth metals with the numbers 61 and 72, the latter of which is perhaps celtium according to Urbain". It was three years later that Coster and Hevesy reported the discovery of hafnium as the 72nd element.

"Furthermore, we now know that, accordingly the Werner form of the periodic table, the rare earth metals are consecutive members of the long cesium period and do not completely fall out of the system. The investigation of the X-ray spectra of the rare earth metals undoubtedly showed that cerium has the atomic number 58, tantalum has the atomic number 73, and that between them the rare earth metals, atomic weights 140.6 to 175.0, fit in with the numbers 59 - 71. The number of theoretically possible rare earth metals is now exactly known."

Concerning the general structure of the periodic table, Pfeiffer wrote "We first have a shortest period (pre-period), consisting of only two elements hydrogen and helium; this is followed by two short periods (Li and Na periods) of eight elements each and two long periods (K and Rb periods) of 18 elements each; Now there is an extremely long period (Cs period), which comprises 32 elements, and a final period (Ra period), from which only individual elements are known so far, so that we cannot say anything about their length. The number of elements of the individual periods can be calculated using the formula $Z = 2n^2$, in which $n$ means the series of integers from 1 to 4."

It is clear that Pfeiffer proposed this new periodic table based on careful assessments of the best scientific knowledge available at the time. Pfeiffer's table preceded better known Deming's table (1923) and is more similar to today's standard table. He later changed the positions of Be and Mg to the group 2 above Ca (Pfeiffer, Fleitmann and Hansen 1930). It is interesting to note that in the biographical articles on Werner (Pfeiffer 1928) and on Pfeiffer (Oesper 1951), there is no mention of their contributions to the development of new types of the periodic tables.

The long form of the periodic tables pioneered by Werner and Pfeiffer have gained much popularity compared with the shot-form table in use in schools and laboratories. On the other hand, the ability to express chemical properties in a compound, such as characteristic oxide formation, becomes weaker in the long-form periodic table. As the group names used until the 1980s, IIA and IIB, for example, express the susceptibility to divalent ions, but the current names of the 2nd and 12th groups have lost such meaning inherited from Mendeleev's law. The long-form periodic table, correctly organizing the electronic configuration of each element, provides perhaps the best arrangement proposed to date. Nevertheless, even this table is not free from some shortcomings. For example, there are unnecessary wide gaps between the $s$ and $p$ blocks in the second and third periods, and similar chemical-valence tendencies among different blocks when forming compounds are not very explicitly represented. In the latter aspect, the long-form periodic table may not be properly called a periodic "law" table in the sense emphasized by Mendeleev.

## 3. 3D periodic table expressing Mendeleev's periodic law: *Elementouch*

In parallel with the proposals of various planar periodic tables, many 3D periodic tables have also been invented. The database by M.R. Leach (Leach 2020) lists about one hundred 3D periodic tables. Among them, "*Elementouch*" (Maeno, 2001) reproduces the periodic character across different blocks of elements as emphasized by Mendeleev.

Mendeleev's periodic law is embodied in his short-form periodic table (Mendeleev 1871). Numerous variations of the short-form table have been proposed (Mazurs 1974, Leach 2020); Fig. 3 shows an example in which we used the same color coding for different blocks of elements as other figures following this. In the 1871 version, the groups forming columns represent the valence properties in forming oxides and hydrides, as clearly indicated in his table as "$R_2O$" and "$RO$" for the groups I and II, "$RH$ and $R_2O_7$" for the group VII (halogens), etc. Such valence tendencies of forming chemical compounds are no longer very explicit in the long-form table based on the electron shell structure. To consider how to bridge between these two types of the tables, let us inspect where different valence tendencies appear in the long-form table based on Janet's left-step periodic table updated as in Fig. 4 (Janet 1929; Scerri 2007). Divalent (less pronounced but also monovalent) tendencies shown in Mendeleev's table as groups II (I) are located in groups 2 as well as 12 (1 as well as 11). Trivalent (tetravalent) tendencies in Mendeleev's groups III (IV) are in groups 3 and 13 (4 and 14), and in addition the first (and the second) columns of the *f*-block elements, La and Ac (Ce and Th).

Fig. 3 Short-form periodic table with the *f*-block elements placed below the main table.

Fig. 4 Left-step periodic table of Janet's style with the old and new group notations at the top. The column locations of the elements with various valence tendencies are indicated at the bottom. The group 18 noble-gas elements are shown with a red frame.

By winding a ribbon of the element symbols in concentric multi-tube helix, it is possible to align elements in group 2 and 12, for example, in the same column and reproduce Mendeleev's arrangements of the group II elements. In order to achieve this, it is necessary to combine *s*- and *p*-block elements in one tube and wind the *d*-block elements in another tube. With the *f*-block elements in the third tube, one can align all trivalent (tetravalent) elements indicated in Fig. 4 in one column. Such division into *three* groups is chemically natural if one recalls that the *s*- and *p*-block elements are collectively called representative elements and provide the framework of the periodic table, the *d*-block elements transition elements, and the *f*-block elements inner transition elements (Allen and Knight 2003, Cao *et al*. 2019).

The 3D periodic table "*Elementouch*" (Maeno 2001, 2002) is constructed by continuously winding a ribbon of element symbols in three-tube helix as shown in Fig. 5. Indeed all divalent, trivalent and tetravalent elements indicated in Fig. 4 line up in the respective columns. In this way it is possible to express the "periodic law" in Mendeleev's short-form periodic table, while keeping the shell structure expressed in the modern long-form table. Mendeleev noted "Cu, Ag, and Au occupy two places – one in the first group (I) and the other in the eighth (VIII)", considering their compounds $Ag_2O$, CuCl and AgCl, and placed them in the group I with parentheses (Mendeleev 1871, Jensen 2002). The divalent state of Cu is well known as the basis of high-temperature superconductivity of cuprates. To express such valence tendensicies, in the *Elementouch*, the group 11 (IB) elements are placed not exactly on the same tube as the group 1 (IA) elements as shown in Fig. 5 (a). This arrangement is also expressed in the pottery model shown below (Fig. 8).

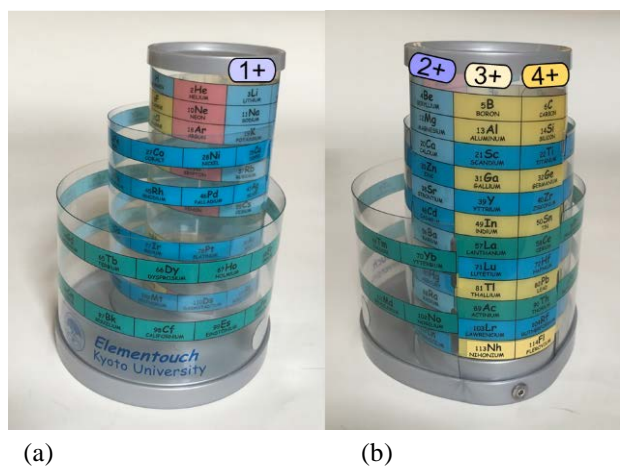

(a)          (b)

Fig. 5 Three-dimensional helical periodic table "*Elementouch*". (a) The three tubes represent *s-p* blocks, *d* block and *f*-block. (b) Divalent, trivalent and tetravalent elements align in the respective columns. Darker blue is used for the *d*-block elements, instead of pink as in Figs. 4 and 9 (b). Shown in these photos is a three-pocket penholder design.

In this winding arrangement, only the trivalent non-magnetic La and Lu among the "rare-earth" elements are arranged in the same vertical column as the group 3 elements such as Y and the group 13 elements such as In (Fig. 5(b)). For long-form 2D periodic tables, it is often argued whether it is La *or* Lu which occupies the same columns as Y (for example Jensen (2018)); in the *Elementouch both* La *and* Lu are in the same column as Y. Nevertheless, it is more natural to treat Lu as the first element in the *d*-block rather than treating it among the "rare earths", since the 4*f* orbitals are filled up with 14 electrons as the other 5*d*-block elements to follow (Landau and Lifshitz 1977). We note that unlike other lanthanides, only La and Lu form non-magnetic ionic states like Y.

Helical periodic tables on concentric multi-tubes have been proposed previously, for example by Schaltenbrand (Schaltenbrand 1920) and Janet (Janet 1928). Starting from Janet's left-step table, it may seem more straightforward to wind the element symbols in four tubes representing *s-*, *p*, *d*, and *f*-blocks. This is indeed what Janet did. Figure 6 compares the top views of the three helical versions discussed here. In Schaltenbrand's helical table with four tubes, the first tube consists of the group 17 and 18 elements plus hydrogen; 1*s* (H and He) and $p^5$ - $p^6$ elements. The second tube contains other *s* and $p^1$ - $p^4$ elements (the groups 1, 2, and 13 - 16). In these tables consisting of four tubes, Mendeleev's periodic law in the sense described above is not explicitly expressed.

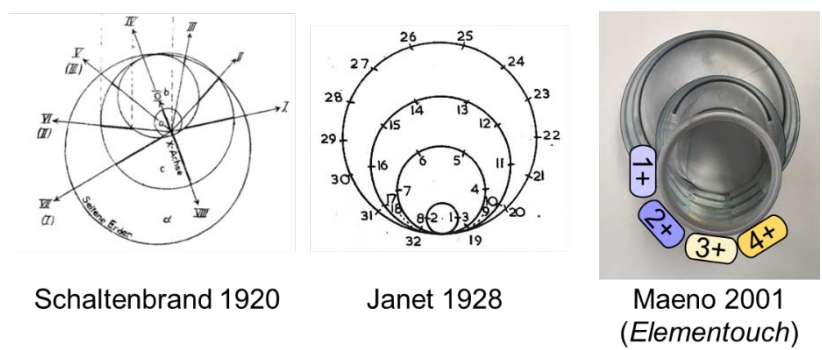

Fig. 6 Comparison of the top views of three concentric helical periodic tables, (a) Schaltenbrand, (b) Janet, and (c) the *Elementouch*.

Mazurs' book classifies numerous periodic tables, including helical tables, many of which were reconstructed or reinterpreted by himself (Mazurs 1957/1974). In his book, a helical table similar to Elementouch is listed as "Vogel 1918" periodic table as shown below in Fig.7 (a): "Helix with three sizes of revolutions for 8 representative, 10 transition, and 14 inner transition elements (Fig. 46). The originator was Vogel in 1918 (table p. 197), who did not draw the helix, but gave only the top view of the helix." However, as shown in Fig. 7 (b) below, the top view of Fig. 2 in Vogel's paper on page 197 is rather different from what Mazurs presented; for example, La and Lu are separated in Vogel's original figure. Thus, this helical table may be better considered as Mazurs's helical table proposed in 1957.

Although the basic arrangements of the elements are the same, there are important differences between Mazurs's table and the Elementouch. In Mazurs table, the three tubes meet on just one line joining six different columns consisting of the groups 3, 13 and 19 on the left and 4, 14, and 20 on the right of the line. In contrast, in the Elementouch the three tubes are fused over the three columns representing divalent, trivalent and tetravelent tendencies, in order to express Mendeleev's law.

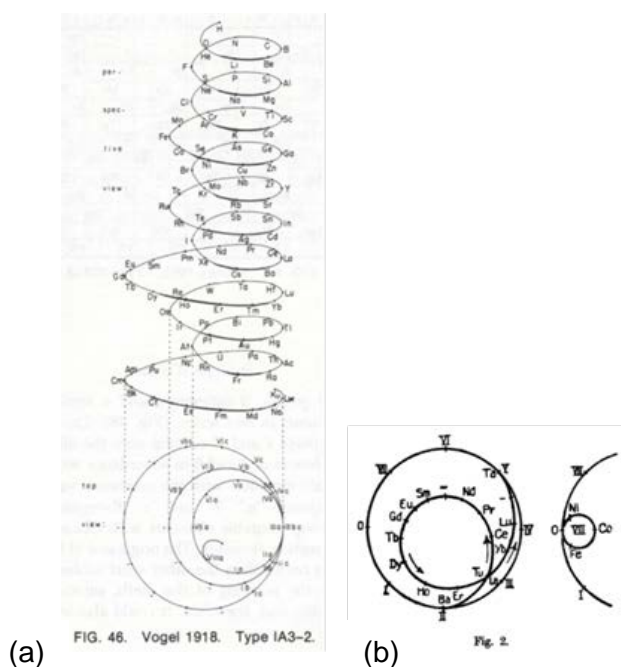

Fig. 7 (a) The three-tube helical table quoted as "Vogel 1918" by Mazurs (Mazurs 1957/1974), compared with (b) the top view of the helical table in Vogel's paper (Vogel 1918).

The multi-tube helical structure of the *Elementouch* is reformed to a simple cylindrical tube, as in the pottery model shown in Fig. 8. To fit the helix on one tube, the width allocation is adjusted; for the group 5 to group 11 elements (Fig. 4) the widths are reduced to 5/7 of the *s*- and *p*-block elements, and for the *f*-block electrons 59 (Pr) - 70 (Yb) and 91 (Pa) - 102 (No), assigned as groups 21 to 32 in Fig. 4, to 6/12. The reduced widths remove unintended matching with Mendeleev's law appearing in the vertical columns. A similar helical design has been adopted as a periodic-table mug cup, as well as to its expanded versions in a towel and a T-shirt (Kyoto University goods).

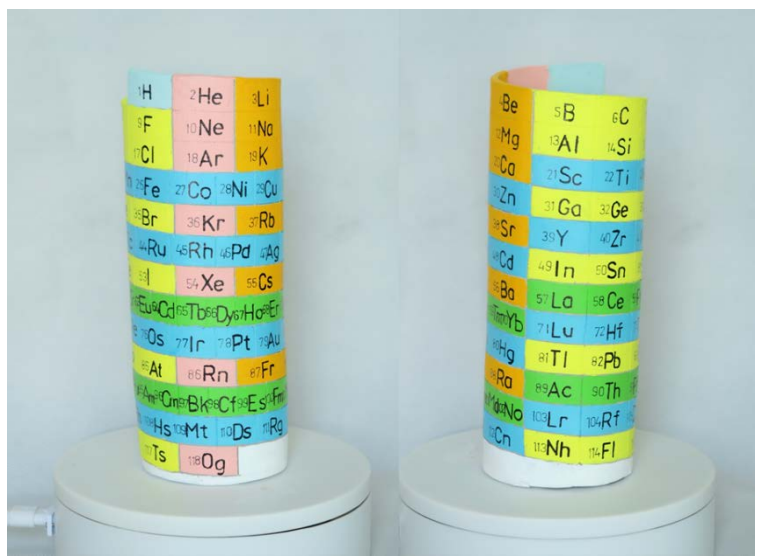

Fig. 8 The rotating pottery model of *Elementouch* on a single helical tube, also available in the video (Ishiguro 2019). Darker blue is used for the *d*-block elements, and green is used for the *f*-block elements, as in Fig. 4. (Left) Noble-gas elements (in pink) and monovalent elements (orange) are aligned in the respective columns; (Right) divalent, trivalent, and tetravalent elements are aligned in the respective columns.

Let us summarize the features of the *Elementouch* compared with the Werner-Pfeiffer planar tables. (1) Elements with similar valence properties are arranged in the same columns, reproducing essential features of Mendeleev's periodic law. (2) Element symbols are lined up seamlessly, without gaps between Be - B and Mg - Al. (3) The *f*-block elements are incorporated continuously and treated equally to the *d*-block elements, instead of being separated as in Pfeiffer's table. (4) From the top, one can depict an atomic model with *s*/*p*, *d*, and *f* orbitals (Fig. 6 (c)). One serious disadvantage in these 3D tables is that one cannot see all the element symbols in one view. To appreciate the arrangements of the helical periodic tables from all directions, a video is available with pottery models (Fig. 8) rotating on a table (Ishiguro 2019).

## 4. Nuclear periodic table and its fortuitous relation to the atomic periodic table

A nuclear periodic table, the "*Nucletouch*", in which elements are arranged based on the proton magic numbers of the nuclear shell structure, has recently been proposed (Hagino and Maeno 2020). As the last topic, we will introduce an additional remark not described in the original paper (Maeno and Hagino 2020): we will extend the comparison between the nuclear and atomic periodic tables shown in Figs. 9 (a) and (b).

Periodicity of the atomic properties of elements originates from the shell structure of the electron orbitals around a nucleus. Under the Coulomb potential, the binding energy of each orbital state is determined by the principal quantum number $n_p = n + \ell$ ($n$: the number of nodes in the radial density, $\ell$: orbital angular quantum number). Except hydrogen with a single electron, interactions with other electrons weaken the binding energies of orbitals with larger values of $\ell$ due to the screening of the nuclear charge especially by the $s$ electrons with $\ell = 0$. This leads to the stability of 4$s$ orbitals over the 3$d$ orbitals, often expressed in terms of the diagram of Madelung's rule (Scerri 2009, 2010). For heavy atoms, spin-orbit interaction originating from relativistic effects becomes important as well.

The long-form periodic table of elements well expresses such shell structure of the electron orbitals. The energy gap between the fully occupied shell configuration and the first excited level acquires maximum values for the group 18 elements, the noble-gas elements, placed on the right most column (Fig. 9 (b)). The chemically inert noble-gas elements are He, Ne, Ar, Kr, Xe (, and Og) with the atomic magic numbers 2, 10, 18, 36, 54, 86 (, and 118). The difference between the neighboring magic numbers is given by $2k^2$ ($k = 2, 3,$ and 4). This is the number of electrons accommodated in each period.

Although a nucleus does not have a potential core at its center, it is well known that protons and neutrons in a nucleus exhibit orbital shell structures similar to those of the electrons in an atom. When orbital shells are completely filled up with protons or neutrons, stable nuclei analogous to noble gas atoms are formed. The magic numbers of protons known for stable nuclei are 2, 8, 20, 28, (40), 50, 82, with a predicted magic number of 114, including a semi-magic number, $Z = 40$. The success of the nuclear shell model was rewarded as the Nobel prize in physics in 1963 to Maria Goeppert Mayer and J. Hans D. Jensen, along with Eugene Wigner.

Figure 9 compares the nuclear and atomic periodic tables of elements. To allow better comparison, the nuclear periodic table here (Fig. 9 (a)) is rearranged to a form similar to Pfeiffer's table; the atomic periodic table (Fig. 9 (b)) highlights proton magic-number nuclei. We also use the notations commonly used for atomic orbitals; $n_pL_j$ ($L = s, p,$ etc. for $\ell = 0, 1,$ etc., and the total angular moment $j = \ell \pm 1/2$). There are major differences in the atomic and nuclear shell structures, as reflected in the difference in the magic numbers. First, instead of the Coulomb potential, the nuclear potential is created by a short-ranged strong interaction among nucleons in the nucleus. The simplest starting point is a 3D harmonic oscillator potential with the cut-off incorporating the range of the nuclear force. The degeneracy of the basic energy levels are set by the parity of the orbitals (even/odd for $\ell$). Notice that as the potential does not have the $1/r$ structure as in the Coulomb potential, the degeneracy feature is significantly different between the Coulomb potential and a harmonic oscillator potential. For instance, the first excited state is a $p$-state in a harmonic oscillator potential while $s$- and $p$- states are degenerate in the Coulomb potential. Second, reflecting the strong spin-dependence of the nuclear force, the spin-orbit interaction is essentially large and plays a much more important role than in atomic systems as the determining factor of the nuclear shell structure. Figure 9 (a) shows that 2$s$ orbitals are in the middle of the spin-orbit split 3$d$ orbitals and that several magic numbers represent the complete filling of only one of the spin-orbit split orbitals. Third, the stability of a nuclide is determined by both the proton and neutron shell structures, whereas an element is distinguished by the number of protons. By choosing the most abundant or most stable nuclide for each atomic number, a meaningful "periodic" table can be constructed: nuclei tend to be spherical and stable near the magic nuclei, and tend to be deformed away from them. These properties are better recognized in the 3D model of the "*Nucletouch*" shown in Fig. 10.

Orbital states expressed in the periodic tables actually contain subtle issues. In the atomic table, the applicability of the Madelung ($n_p + l, n_p$) rule is rather involved (Allen and Knight 2003, Schwarz 2010, Cao *et al.* 2019, Scerri 2020). Stronger core-charge screening by the $s$ electrons, increasing $n_p$ energy splitting with increasing $Z$, and stronger electron correlation (Coulomb repulsion) among the more compact $d$ electrons compared with $s$ electrons, etc. lead to complications. In fact, there are a number of elements with irregular order of electron occupancies: eleven elements in the $d$-block including Lr (such as Cr with $4s^1 3d^5$, Pd with $4s^0 3d^{10}$, and Lr with $4s^2 3p^1$) and nine elements among the $f$- block (such as La with $6s^25d^1$ and Th with $7s^26d^2$ without any $f$ electron).

Additional consideration for the nuclear orbitals is that if the nucleus is deformed from the spherical shape, a proton is in a hybridized state consisting of different single-particle orbital states. Nevertheless, it is worth assigning a single-particle orbital occupancy for each element to illuminate the comparison between the atomic and nuclear periodic tables (Figs. 9 (a) and (b)). Indeed, it is interesting to find unexpectedly that the magic nuclei Sn-Pb-Fl are aligned in the same column also in the ordinary "atomic" periodic table. This is because the increment of electron numbers 32 of the noble-gas elements from Xe (54) to Rn (86) and to Og (118) is identical to the increment of proton numbers among the magic nuclei Sn (50), Pb (82) and Fl (114). Thanks to this coincidence, the alignments of the elements in both periodic tables are very similar after Nb (41). Ag (47) is just above Au (79) and La (57) is just above Ac (89), as examples.

Fig. 9 (a) Nuclear periodic table "*Nucletouch*" (Hagino and Maeno 2020), (b) atomic periodic table with the proton magic nuclei highlighted with bold character, and (c) a schematic to compare the atomic and nuclear shell structures.

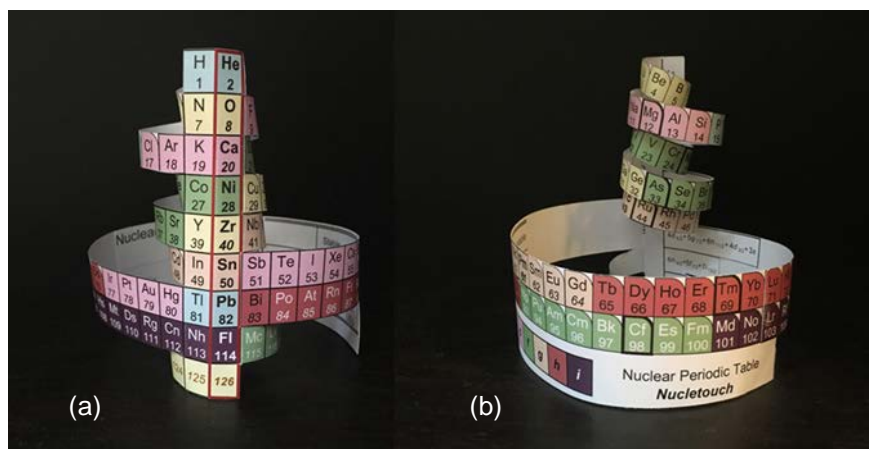

Fig. 10 Three dimensional model of the nuclear periodic table "*Nucletouch*" with the colors representing single-orbital states. (a) Front view: proton magic number elements are aligned instead of noble-gas elements. (b) Rear view: elements with maximum nuclear deformation are shown in black frames.

## 5. Conclusions

We introduced three related topics on the periodic tables of elements. As a pioneer of the long-form periodic table, Werner's contribution is worth more proper recognition among wider community, especially through high-school textbooks. Pfeiffer's contribution to the modern arrangement of the long-form table seems much less known even among the specialists in the field. We would like to emphasize that Pfeiffer introduced the arrangement of the *f*-blocks as a separate table based on his profound knowledge from the X-ray physics.

As the second topic, we explained how the 3D helical periodic table well reproduces the essence of Mendeleev's periodic law, by winding into three tubes with the *s*- and *p*-block elements combined in one tube. Elements with the same ionic tendency are aligned in the same column, unlike some other 3D helical tables proposed. The "periodic law", such as the similarity between the groups 2 (IIA) and 12 (IIB) elements, is clear in the *"Elementouch"* periodic table. Beyond Mendeleev's law, valance similarities among the groups 3 (IIIA) and 13 (IIIB), as well as among the groups 4 (IVA) and 14 (IVB), are extended to include some of the relevant *f*-block elements. Thus, both La and Lu are in the same column as Y in the *Elementouch*.

The nuclear periodic table *"Nucletouch"* may be expanded if the neutron numbers can somehow be incorporated in a concise form, but this is a future issue. The introduction of the nuclear periodic table implies that there may be yet some other forms of "periodic" tables that represent other distinct properties of elements. For instance, production cross sections of superheavy nuclei are enhanced for several reasons when nuclei with magic numbers and their neighbors are used in experiments, and the nuclear periodic table may be helpful in understanding why Ca, Pb and Bi have been used to synthesize superheavy elements up to Og (Hamilton, Hofmann, and Oganessian 2013; Hagino 2019).

Finally, we mention that a variety of patterns to make models of *Elementouch* and *Nucletouch* can be downloaded from: *http://www.ss.scphys.kyoto-u.ac.jp/elementouch/index.html*

**Acknowledgements:** We acknowledge Yoji Hisamatsu for the valuable information of Pfeiffer's original literature, Kohei Tamao, Kaoru Yamanouchi, and Nagayasu Nawa for important information, and Markus Kriener and Robert Peters for help in translating the original papers. We appreciate the reviewers of this paper for their critical comments and very useful suggestions. YM acknowledges supports by Japan Society for the Promotion of Science (JSPS) Core-to-Core Program (A. Advanced Research Network) and JSPS KAKENHI Grant No. JP17H06136.